# A new method to scan genomes for introgression in a secondary contact model

ANTHONY J. GENEVA, SARAH B. KINGAN, and DANIEL GARRIGAN

*Department of Biology, University of Rochester, Rochester, New York 14627 USA*

Correspondence: Daniel Garrigan, Fax +1 585-275-2070
Email: dgarriga@ur.rochester.edu

**Abstract**

**Secondary contact between divergent populations or incipient species may result in the exchange and introgression of genomic material. We develop a simple DNA sequence measure, called $G_{min}$, which is designed to identify genomic regions experiencing introgression in a secondary contact model. $G_{min}$ is defined as the ratio of the minimum between-population number of nucleotide differences to the average number of between-population differences. One advantage of $G_{min}$ is that it is computationally inexpensive relative to model-based methods for detecting gene flow and it scales easily to the level of whole-genome analysis. We compare the sensitivity and specificity of $G_{min}$ to those of the widely used index of population differentiation, $F_{ST}$, and suggest a simple statistical test for identifying genomic outliers. Extensive computer simulations demonstrate that $G_{min}$ has both greater sensitivity and specificity for detecting recent introgression than does $F_{ST}$. Furthermore, we find that the sensitivity of $G_{min}$ is robust with respect to both the population mutation and recombination rates. Finally, a scan of $G_{min}$ across the X chromosome of *Drosophila melanogaster* identifies candidate regions of introgression between sub-Saharan African and cosmopolitan populations that were previously missed by other methods. These results show that $G_{min}$ is a biologically straightforward, yet powerful, alternative to $F_{ST}$, as well as to more computationally intensive model-based methods for detecting gene flow.**

**Introduction**

Secondary contact occurs when sympatry is restored between two or more populations that have evolved for some amount of time in allopatry. For evolutionary biologists, secondary contact between diverging populations can provide a compelling natural experiment. For example, the frequency and symmetry of hybrid matings can yield insight into the roles of sexual selection (Ritchie 2007) and/or reinforcement (Yukilevich 2012) in speciation. Likewise, the frequency of backcrossing and subsequent introgression can reveal the extent to which postzygotic isolating mechanisms have accumulated (Gompert *et al.* 2012). In this context, studies of naturally occurring secondary contact offer a distinct advantage over laboratory-based studies of reproductive isolation— the patterns of introgression represent the fitness of hybrid genotypes in natural environments, replete with a variety of ecological selection pressures. Lastly, studies of secondary contact are not limited merely to satisfying the intellectual curiosity of evolutionary biologists: hybridization and introgression from closely related invasive populations can be a significant extinction threat for endangered endemic populations (Rhymer & Simberloff 1996; Seehausen *et al.* 2008).

With the advent of comparative population genomics, there is now the potential to quantify the frequency and tempo of genomic introgression between natural populations experiencing secondary contact. While these studies are just beginning to appear, many investigators rely upon low values of the traditional fixation index, $F_{ST}$ (Wright 1951), to identify introgressing genomic regions (*e.g.*, Nadachowska-Brzyska *et al.* 2013; Neafsey *et al.* 2010; Smith & Kronforst 2013). We suggest that $F_{ST}$ may not be ideally suited for this particular application: it is derived from the variance in allele frequencies among populations and may lack power to detect introgression in cases of secondary contact (Murray & Hare 2006). This is because for $F_{ST}$ to take on values close to zero following secondary contact, alleles must not only be shared across populations, and their frequencies in the two populations must be equal. This is not necessarily expected in a secondary contact model in which introgression is either very recent or otherwise limited. In this paper, we consider whether whole-genome sequence data can be leveraged to obtain both greater sensitivity and specificity to detect introgression than using $F_{ST}$ alone. While there are a variety of alternatives to $F_{ST}$ for detecting introgression (Barton *et al.* 2013; Gompert & Buerkle 2011; Harris & Nielsen 2013; Machado *et al.* 2002; Pool *et al.* 2012; Price *et al.* 2009; Ralph & Coop 2013; Sousa & Hey 2013), our aim is to develop a method that uniquely fulfills seven criteria: 1) it has minimal prior assumptions, 2) is sensitive to recent gene flow, 3) has a low rate of false positives, 4) has a straightforward biological interpretation, 5) is applicable to a wide range of taxa, 6) can localize tracts of introgression in the genome, and 7) is fast to compute on large genomic datasets. To this end, we propose a simple haplotype-based sequence measure called $G_{min}$, which is can be quickly calculated in a sliding window across whole-genome alignments. $G_{min}$ is the ratio of the minimum between-population haplotype distance to the mean between-population haplotype distance, calculated in windows across the genome. We present the results of extensive computer simulations demonstrating that $G_{min}$ is more sensitive to recent introgression than $F_{ST}$ in a secondary contact model. We also use $G_{min}$ on a previously published dataset to scan the X chromosome for introgression between sub-Saharan African and cosmopolitan populations of the commensal fruit fly *Drosophila melanogaster.*

**Materials and Methods**

*Rationale for the $G_{min}$ measure*

Assume that we have nucleotide sequences of multiple individuals sampled from two populations, such that there are a total $n_1$ sequences from population 1 and $n_2$ sequences from population 2. The average number of pairwise nucleotide differences between sequences from the two populations is defined as

$$\bar{d}_{xy} = \frac{1}{n_1 n_2} \sum_{x=1}^{n_1} \sum_{y=1}^{n_2} d_{xy}, \qquad (1)$$

in which $d_{xy}$ is the Hamming distance (or, *p*-distance) between sequence *x* from population 1 and sequence *y* from population 2 (Nei & Li 1979). Similarly, let $\min(d_{xy})$ be the minimum value of $d_{xy}$ among all $n_1 \times n_2$ comparisons. We can then define the ratio,

$$G_{min} = \frac{\min(d_{xy})}{\bar{d}_{xy}} \qquad (2)$$

The ratio $G_{min}$ ranges from zero to unity and has the property that if $n_1 = 1$ and $n_2 = 1$, then $G_{min} = 1$. Under a strict model of isolation (*i.e.*, no historical gene flow), a lower bound is imposed upon $G_{min}$ by the divergence time between the two populations. However, no such lower bound exists for population divergence models that include recent gene flow (for example, see **Fig. 1**).

*Behavior of the $G_{min}$ ratio*

To characterize the behavior of the $G_{min}$ ratio, two sets of coalescent simulations were generated. The first set was intended only to examine the distribution of $G_{min}$ under the null model of neutral population divergence with no gene flow (isolation). The second set of simulations was designed to contrast the sensitivity and specificity of $G_{min}$ with those of $F_{ST}$, using a binary classification procedure. This second set considers a large parameter space for a secondary contact model, which includes an ancestral population of size *N* that splits into two descendant populations at time $\tau_D$ (measured in units of *N* generations). We focus on cases in which each of the descendant populations also has size *N* (however, for treatment of the effects of varying population size in secondary contact models, see Geneva & Garrigan 2010). Subsequently, at time $\tau_M$ (also measured in units of *N* generations) before the present, the source population is allowed to send migrants instantaneously to the other population. Instantaneous migration was assumed, rather than specifying a time for the onset of continuous gene flow, because it more discretely captures the effect of the timing of secondary contact. The number of migrating lineages is governed by the "migration probability" parameter, $\lambda$. For example, at time $\tau_M$, let there be $k$ ancestral lineages present in the source population, so that the number of lineages chosen to migrate is a binomial distributed random variable with expectation $k\lambda$. We assume that gene flow is unidirectional. This model is implemented in a modified version of the coalescent simulation software MS (Hudson 2002), called MSMOVE (Garrigan & Geneva 2014).

Since $G_{min}$ is intended to be measured in a sliding window scan of whole-genome sequence alignments, we performed simulations that approximate variably sized genomic windows. This was achieved by varying both the population mutation rate ($\theta = 4N\mu$ where $\mu$ is the mutation rate for a given window) and the population crossing-over rate ($\rho = 4Nc$, where $c$ is the rate of crossing-over per window). Specifically, we used values of $\theta \in \{10, 20, 50, 100, 150\}$ and $\rho \in \{0, 1, 10, 20, 50, 100, 150\}$. For a sample size of 10 individuals, it is expected that $\theta = 10$ corresponds to a window size with 28 segregating sites, while $\theta = 150$ approximates a window with 424 segregating sites. Similarly, $\rho$ approximates the size of haplotypes within windows. For example, when $\theta = 150$ and $\rho = 0$, all 424 segregating sites would be partitioned among haplotypes that span the length of the window. However, when $\theta = 150$ and $\rho = 150$, there are also 424 expected recombination events, therefore each segregating site would have its own non-recombining coalescent history, on average.

For each pairwise combination of parameter values, a total of $10^4$ independent windows were simulated. This scheme assumes that large windows are being used to scan the genome for gene flow, such that genealogical histories within windows can be correlated, but that adjacent windows contain independent genealogies. Additionally, we considered two different sample size configurations. The first configuration is one in which only a single source-population sequence is available ($n_1 = 10$ and $n_2 = 1$) and the second sample configuration assumes that polymorphism data are available from both populations ($n_1 = 10$ and $n_2 = 10$). For both sample size configurations, the direction of the gene flow is from population 2 into population 1, going forward in time.

For the first set of simulations, which characterizes the behavior of $G_{min}$ under the null isolation model, we considered a range of population divergence times, $\tau_D \in \{1/25, 2/25, 3/25, \ldots, 8\}$. We performed a variance partitioning analysis to quantify the effects of the $n_2$, $\theta$, $\rho$, and $\tau$ parameters (as well as their interactions) on the mean and variance of both $G_{min}$ and $F_{ST}$. We first fit a linear model that includes all parameters and their interactions. We then quantified the variance explained by each parameter by calculating the partitioned sum of squares. For all analyses, we tested the non-independence of parameters and for any potential bias-inducing effects of model complexity by comparing variance partitioning for each parameter after 1) iterating the order of parameters in the model, 2) running models both with and without interaction terms, and 3) serially removing parameters. All post-processing and analyses of simulated data was performed using the R statistical environment (R Core Team 2013).

*Sensitivity and specificity*

To contrast the sensitivities of $G_{min}$ and $F_{ST}$ to gene flow under the alternative secondary contact model, we examine the proportion of simulated true migrant genealogies that appear in the tails of the distributions. While this is not meant to be a formal statistical test, it is a convenient procedure for approximating the sensitivity and specificity of $G_{min}$ and $F_{ST}$. Using this procedure, we classify a genomic window as being "positive" for gene flow on the basis of a one-tailed $Z$-test, which considers the area under the negative tail of the $Z$-scores. To identify outliers, we assumed the $Z$-scores are normally distributed and we consider three different $P$-value thresholds for positive classification: 0.05, 0.01, and 0.001. Let the set of windows with a $P$-value less than the threshold be denoted as $Q$. Furthermore, simulated windows are classified as "true" gene flow windows if they contain a genealogy in which an ancestral lineage has switched populations. Therefore, any particular parameterization of the secondary contact model will yield the set $M$ of true gene flow windows. Let $M \cap Q$ represent the set of true gene flow windows with a $P$-value below the threshold value. The sensitivity of the test ($\varphi$) can therefore be defined as the proportion

$$\varphi = \frac{|M \cap Q|}{|M|} \tag{3}$$

Thus, $\varphi = 1$, when all true gene flow windows have a significant $P$-value. Conversely, we define specificity ($\psi$) as

$$\psi = \frac{|M \cap Q|}{|Q|}, \tag{4}$$

such that if $\psi = 1$, then all windows with a significant $P$-value are true gene flow windows. For the analysis of sensitivity and specificity, the simulated parameter combinations were the same as those used in the first set of simulations described in the previous subsection. The only exceptions were that we simulated a narrower range of divergence times $\tau_D \in \{1/100, 2/100, 3/100, \ldots, 1\}$ and added two additional parameters: the relative time of gene flow, which had the range $\tau_M \in \{\tau_D/100, 2\tau_D/100, 3\tau_D/100, \ldots, \tau_D\}$ (for $\tau_D > 0$) and migration probability in the set, $\lambda \in \{0.001, 0.005, 0.01, 0.05, 0.1\}$. In addition to assessing the sensitivity and specificity of $G_{min}$ and $F_{ST}$, we also evaluated the effect of each varied parameter on sensitivity and specificity. Variance partitioning was performed as described in the previous subsection.

*Application to* Drosophila melanogaster *data*

We developed $G_{min}$ in anticipation of high-quality short-read assemblies of population-level samples from more than one population. Such data have just begun to emerge from a variety of organisms. To contrast the sensitivity of $G_{min}$ with that of $F_{ST}$, we apply it to a subset of the highest quality available resequence dataset: X chromosome polymorphism of two populations of *Drosophila melanogaster* (Pool *et al.* 2012). The two populations include a cosmopolitan population from France and a sub-Saharan African population from Rwanda. While these two populations generally show low levels of sequence divergence (chromosome average $F_{ST} = 0.183$ and $\bar{d}_{xy} = 0.0085$), a recent study was able to detect a signal of recent cosmopolitan admixture in several African populations, including the deeply sampled Rwandan population (Pool *et al.* 2012). We obtained 76 bp paired-end Illumina reads from seven French and nine Rwandan lines from the NCBI short read archive (see **Table S1** for details on the sampled lines). All reads were aligned to the reference genome of *D. melanogaster*, build 5.45

(http://flybase.org), using the BWA software, version 0.6.2 (Li & Durbin 2009). The resulting alignments for individual lines in the BAM format were merged using the SAMTOOLS software package (Li *et al.* 2009). The values of $F_{ST}$ and $G_{min}$ were calculated in non-overlapping 50 kb windows using the POPBAM software package (Garrigan 2013). We only analyzed nucleotide sites that met the following criteria: read depth per line greater than 5, Phred-scaled scores for the minimum root-mean squared mapping quality greater than or equal to 25, and a SNP quality that is at least 25; we also only incorporated reads with a minimum mapping quality of 20 and an individual base quality of at least 13. Genomic windows with less than 25% of the reference genome positions passing the above filters were subsequently ignored. Lastly, we construct neighbor-joining trees based on uncorrected Hamming distance in 50 kb windows using POPBAM. Individual windows were identified as outliers using Z-test described above with a significance threshold of $P < 0.05$ for $G_{min}$. We compare our analysis to that of Pool *et al.* (2012), who utilized a Hidden Markov Model method based on the pairwise distances between sub-Saharan African and cosmopolitan genomes. In windows of 1000 non-singleton SNPs, each Rwandan line was assigned a posterior probability of admixture. We identified previously known admixed regions as those whose sum of posterior probabilities across lines is greater than 0.50 (see Table S5 from Pool *et al.* 2012). Finally, it should be noted that we selected a subset of the French and Rwandan lines from the original study with the highest genome coverage reported by Pool *et al.* (2012).

**Results**

*Behavior of the $G_{min}$ statistic under an isolation model*

$G_{min}$ is the ratio of min($d_{xy}$), the minimum number of nucleotide differences between haplotypes sampled from different populations, to $\bar{d}_{xy}$, the average number of between-population differences (**Eq. 2**). In a strict isolation model of divergence, we expect that both min($d_{xy}$) and $\bar{d}_{xy}$ will increase as a function of the population divergence time, $\tau_D$. Ultimately, $G_{min}$ is expected to approach unity for very ancient divergence times ($\tau_D \gg 4N$), because there is a high probability of only a single ancestral lineage remaining in each population. Conversely, for very recent divergence times, $G_{min}$ is expected to be much less than unity, since it is unlikely that all coalescent events will occur only between ancestral lineages from the same population before a single coalescent occurs between lineages from different populations. Computer simulations show that both $G_{min}$ and $F_{ST}$ increase asymptotically to unity as the divergence time increases, but also that $G_{min}$ increases at a faster rate and plateaus at an earlier divergence time (**Fig. 2**).

In the isolation model, the variance of $G_{min}$ is most strongly affected by the time of population divergence, $\tau_D$. Variation in $\tau_D$ alone explains approximately half of the simulated variance for both $G_{min}$ and $F_{ST}$ (**Table 1**). When the population mutation rate $\theta \leq 10$, $G_{min}$ becomes downwardly biased (**Fig. 2B**). We suspect that this bias arises for low mutation rates because, when few mutations occur on a set of correlated genealogies, $G_{min}$ does not always capture the minimum time of the between-population coalescent events, rather it may reflect a randomly chosen between-population coalescent event that, by chance, has fewer mutations separating them than the true minimum event. Finally, whether a single source-population sequence is available ($n_2 = 1$) or polymorphism data are available ($n_2 = 10$) has a minor, but predictable, effect: $G_{min}$ is always closer to unity when $n_2 = 1$ than when $n_2 = 10$ (**Fig. 2**). It should be noted that although we report on the results for $F_{ST}$ in the case of $n_2 = 1$, this is obviously not a situation in which $F_{ST}$ (as a measure of difference in allele frequencies) would be applicable. Finally, we found no evidence of bias in any of the variance partitioning analyses, so that the full models with all parameters and interaction terms have been included.

*Sensitivity and specificity*

When we consider a secondary contact model, the two parameters that exert the strongest influence on the behavior of both $G_{min}$ and $F_{ST}$ are the time of migration relative to divergence ($\tau_M$) and the magnitude of the migration event ($\lambda$) (**Tables S2** and **S3**). Our simulations show that $G_{min}$ has increased sensitivity and specificity compared to $F_{ST}$ for all combinations of the $\tau_M$ and $\lambda$ parameters, regardless of the values of nuisance parameters such as $\theta$ and $\rho$ (**Fig. 3**). The sensitivity of $G_{min}$ is greatest when $\tau_M$ is recent and $\lambda$ is small (**Fig. S1**). It is interesting to note that the sensitivity of $G_{min}$ decreases with increasing $\lambda$ because large amounts of migration tends to reduce the average between-population sequence distance, thereby also reducing the expected $G_{min}$ and increasing its variance (**Table S4**). However, for $F_{ST}$, $\lambda$ does not have a profound effect on its sensitivity (**Table S4**). In contrast, increased $\lambda$ results in a greater specificity for $G_{min}$ (**Fig. 3**). This means that although high $\lambda$ results in a lower proportion of the migrant genealogies appearing in the negative $Z$-score tail, a greater proportion of all genealogies in the tail are true migrant genealogies.

Surprisingly, the rate of recombination has only a mild effect on the sensitivity of $G_{min}$ and $F_{ST}$ (**Fig. S2**). This may be due to the relatively intermediate levels of recombination used in the computer simulations, since the recombination rate must be very high ($\rho > 50$) to break up introgressed haplotypes when $\tau_M$ is very recent. This is also true of specificity (**Fig. S3**). Likewise, increasing the population mutation rate also slightly increases both the sensitivity (**Fig. S4**) and the specificity (**Fig. S5**). These results suggest that sensitivity and specificity of $G_{min}$ are optimal when large genomic windows ($\theta > 10$) with relatively low levels of recombination ($\rho < 20$) are considered.

A trade-off between sensitivity and specificity occurs when we contrast results from simulations of divergence from a single source population sequence ($n_2 = 1$) with those from polymorphism data from both populations ($n_2 = 10$). $G_{min}$ has increased sensitivity when $n_2 = 1$ compared to when $n_2 = 10$ (**Fig. S6**). In contrast, the specificity of $G_{min}$ is substantially greater when $n_2 = 10$ (**Fig. S7**). Therefore, situations in which only a single source-population sequence is available results in $G_{min}$ having increased power to detect migrant genealogies at any given locus in the genome, while polymorphism data from two populations yields increased power to detect gene flow across the genome. The specificity result is intuitive from a biological standpoint: if low levels of gene flow occur, then having more sequences per population will increase the probability of recovering an introgressed haplotype. Sensitivity increases when $n_2 = 1$ because there is less variance in the coalescent process in the ancestral population for genealogies that do not experience gene flow and the expected $G_{min}$ in an isolation model is closer to unity;

this results in a higher proportion of migrant genealogies significantly departing from a genome-wide distribution.

*Application to cosmopolitan admixture in* Drosophila melanogaster

We compare the ability of $G_{min}$ *versus* $F_{ST}$ to detect cosmopolitan admixture in a Rwandan population of *D. melanogaster*. We calculated the two statistics in 436 non-overlapping 50 kb windows on the X chromosome in a sample of seven French and nine Rwandan lines (**Fig. 4A**). The mean and standard error for $G_{min}$ is 0.6500 ± 0.0311 and for $F_{ST}$ is 0.1725 ± 0.0083. Interestingly, the range of $G_{min}$ (0.0982-0.9833) is more than twice as large as that of $F_{ST}$ (0.0170-0.5107) (**Fig. 4B**). This expanded range of $G_{min}$ is consistent with a greater sensitivity for $G_{min}$, even for relatively low levels of population divergence. The outliers from the chromosome-wide $G_{min}$ distribution identified cosmopolitan admixture in all of the previously identified admixture windows (**Fig. 4A**). In contrast, outlier values of $F_{ST}$ appear in only one of the six previously identified tracts (**Fig. 4A**). The outliers of $G_{min}$ also reveal two additional candidate introgression tracts on the X chromosome— a region consisting of five significant windows between coordinates 1.65-2.05 Mb, and a single marginally significant ($P = 0.051$) window located at 12.95-13 Mb; neither region was previously identified by Pool *et al.* (2012). The first region near the 2 Mb coordinate harbors a low frequency introgressed haplotype carried by Rwandan line, RG35. Neighbor-joining trees indicate that the RG35 sample is nested within the French samples, although the particular French line(s) with which it clusters varies across windows (**Fig. S8**). The second marginally significant window involves a similar scenario where RG35 is nested within the clade of French lines, sister to the French line FR229 (**Fig. S9**). These inferred low frequency introgressions went undetected in both our $F_{ST}$ scan and the Hidden Markov Model analysis performed by Pool *et al.* (2012). The window size used by Pool *et al.* (2012) was based on the number of SNPs, rather than physical distance, such that windows in this sub-telomeric region are larger than 100 kb, on average. Therefore, it is possible that the large windows analyzed by Pool *et al.* (2012) contain conflicting genealogical histories, resulting in the distance between RG35 and any particular French line not being reduced, on average.

**Discussion**

Comparative population genomic datasets, or whole genome alignments of many individuals from multiple populations within a species or between closely related species, are finally becoming realized in evolutionary genetics. One of the many potential uses of these new data is to estimate the degree to which introgression occurs in populations coming into secondary contact. Also of interest is identifying the genomic location of introgression and the functional properties of introgressing coding material, if any. Many of the first studies to make use of whole-genome datasets rely on the traditional fixation index, $F_{ST}$, to identify introgressed genomic regions. However, $F_{ST}$ has a number of inherent weaknesses for detecting introgression in a secondary contact model.
Phased haplotype data can be especially useful for inferring aspects of historical population demography and gene flow (Machado *et al.* 2002; Pool *et al.* 2010; Pool & Nielsen 2009) and haplotype sharing among populations is often used as a criterion for detecting introgression (Hufford *et al.* 2013; Kijas *et al.* 2012; Ralph & Coop 2013). We show that haplotype-based measures of within- and between-population sequence differences, such as $G_{min}$, offer better sensitivity and specificity over allele frequency measures such as $F_{ST}$. Furthermore, our simulations also show that $G_{min}$ is robust to local variation in mutation rate and, to a lesser extent, recombination rate. The robustness of $G_{min}$ to the local recombination rate primarily occurs when gene flow is both recent and limited, in which case there is a limited opportunity for recombination to break up introgressed haplotypes (**Figs. S2** and **S3**). This result suggests that choice of window size offers an avenue for distinguishing recent versus older introgression events (**Fig. S10**). Larger windows with more mutation and recombination events are more likely to identify very recent introgression events, whereas smaller windows can identify older introgression events, albeit with less specificity than larger windows.

Like $F_{ST}$ or $\bar{d}_{xy}$, $G_{min}$ is not a formal test statistic, rather it is sequence measure designed to identify a distinctly bimodal pattern of between-population coalescence that is expected under models of secondary contact, but not expected in models of strict population isolation. We were unable to derive a closed-form expression for the variance of the $G_{min}$ ratio in a pure isolation model, due in part to the fact that we observe a non-zero positive covariance between the numerator, $\min(d_{xy})$, and the denominator, $\bar{d}_{xy}$ (data not shown). Therefore, using $G_{min}$ as the basis for a simple single-locus test is not currently feasible. However, like $F_{ST}$, $G_{min}$ can be readily incorporated into other inferential frameworks, such as approximate likelihood methods (Beaumont *et al.* 2002). Our approach differs from more formal inferential frameworks, such as those used by the IM program (Sousa & Hey 2013), in that IM tests the hypothesis of whether or not gene flow has occurred; the goal of $G_{min}$ is less formal, seeking to localize introgression genealogies in otherwise diverging genomes. In practice, a $G_{min}$ scan may be an extremely useful first step for identifying candidate regions for introgression. Unlike many likelihood-based methods for detecting gene flow in a population divergence model, $G_{min}$ can be quickly applied to large whole-genome datasets and interpretation of $G_{min}$ requires a minimal set of assumptions. The fundamental assumption is that the individuals in the analysis came from either one population or a different population. This is in contrast to some methods for detecting admixed regions of the genome, which rely on investigators being able to assign individuals to two pure parental populations, as well as a third population of hybrid individuals (Price *et al.* 2009). Of course, knowing the hybrid status of individuals, or having more detailed information of sample geographical distribution, may enable more advanced analysis (Barton *et al.* 2013; Gompert & Buerkle 2011).

While $G_{min}$ is more sensitive to recent gene flow than $F_{ST}$, it has additional desirable properties that distinguish it from other recently proposed haplotype-based methods. For example, Harris and Nielsen (2013) describe a method for detecting recent gene flow by measuring the genomic length distribution of tracts of identity-by-state. The computer simulations presented by Harris and Nielsen (2013) demonstrate that their method can accurately infer the timing and magnitude of admixture events, as well as other demographic parameters, over a range of time scales. However, the identity-by-state method of Harris and Nielsen (2013) may also be sensitive to 1) low quality reads and sequencing error, 2) reductions in effective population size due to background selection, and 3) historical population bottlenecks. In contrast, we argue that $G_{min}$ is not as sensitive to errors in sequencing or assembly, because $G_{min}$ does not explicitly depend upon uninterrupted runs of shared polymorphic sites. Additionally, the lower tail of $G_{min}$ is not expected to be strongly affected by background selection under a secondary contact model. This is because background selection does not affect the tempo of neutral divergence (Birky & Walsh 1988) and can skew within-population polymorphism towards an excess of rare alleles (Charlesworth *et al.* 1995), neither of which affects $G_{min}$ (however, for the effect of reductions in the effective population size, see below). Besides recent introgression, the primary factor affecting $G_{min}$ is the number of ancestral lineages present at the time of the initial population split. As a result, the distribution of $G_{min}$ will be affected by any force that alters the probability density of within-population coalescent events, including changes in the effective population size or natural selection. If natural selection acts to reduce diversity in one population exclusively or, if the effective population size of one population is smaller than that of the other, we expect there to be fewer ancestral lineages present at the time of the initial population divergence. To consider the performance of $G_{min}$ in these cases, we can extrapolate from our computer simulation results of different sampling schemes, in particular when $n_1 = 10$ and $n_2 = 1$. We find that when only a single source-population genome is used, $G_{min}$ has greater sensitivity (**Fig. S6**), but reduced specificity compared to when $n_2 = 10$ (**Fig. S7**). This suggests that forces acting to increase the rate of coalescence within populations, such as population bottlenecks, will result in increased confidence that small values of $G_{min}$ can be attributed to recent gene flow, but also a diminished ability to recover all of the introgressed regions in a genome. Similarly, the reduced specificity of $F_{ST}$ when there is a reduction in within-population variation is well-known (Charlesworth 1998; Cruickshank & Hahn 2014; Nei 1973), however $G_{min}$ does not appear to be as strongly affected as $F_{ST}$ (**Fig. S7**).

In conclusion, we do not wish to argue that $G_{min}$ is in any way a panacea for the longstanding problem of distinguishing models of gene flow from those of pure isolation (Takahata & Slatkin 1990). Indeed, $G_{min}$ lacks sensitivity when gene flow occurred more than halfway back to the time of the population divergence or when there is a large amount of gene flow (**Fig. S1**). For example, if a genomic region is sweeping across species boundaries (Brand *et al.* 2013), $G_{min}$ is not expected to be as informative as $F_{ST}$. Therefore, it is also important to caution that genomic intervals with low $G_{min}$ should be subsequently vetted to ensure that the region does not have unusually low values of absolute $\bar{d}_{xy}$. However, in cases of recent secondary contact, and when the rates of gene flow are not extremely high, we have shown that $G_{min}$ performs well and is more reliable than $F_{ST}$ (**Fig. 3**). In addition, we illustrate how a simple statistical procedure employing $G_{min}$ to scan the X chromosome of recently diverged cosmopolitan and sub-Saharan African populations of *Drosophila melanogaster* performs as well as more sophisticated methods (**Fig. 4**). However, unlike many more sophisticated methods, the calculation of $G_{min}$ is fast and broadly applicable to any taxa for which haploid genome sequences are available. $G_{min}$ can be easily calculated from population genomic data using the software package POPBAM (Garrigan 2013). We anticipate that with the continued emergence of new haplotype sequencing methods (Kirkness *et al.* 2013; Langley *et al.* 2011), these types of data will be increasingly used for evolutionary studies. In this case, $G_{min}$ can be an effective and biologically straightforward addition to the suite of tools available to evolutionary biologists.

## Acknowledgements

We would like to thank Sohini Ramachandran, Jeff Wall, Thomas Mailund, Carlos Machado, and an anonymous reviewer for valuable comments on earlier drafts of this manuscript. We also thank Christina Muirhead and LeAnne Lovato for preliminary work on this project. This work was made possible by a grant from the National Institutes of Health to DG (R01-ODO1054801).

## Author contributions

A.J.G. and D.G. conceived the research, designed the study, and wrote the paper. A.J.G. did the coalescent simulations. S.B.K. performed the analysis of the *Drosophila melanogaster* data and contributed to writing the paper.

**Data Accessibility**
Summary statistics calculated from coalescent simulations. Dryad doi:XXXXXXX
**Supporting information**
**Table S1.** The sampled lines from two populations of *Drosophila melanogaster*.
**Table S2.** Analysis of variance of sensitivity of $G_{min}$ and $F_{ST}$.
**Table S3.** Analysis of variance of specificity of $G_{min}$ and $F_{ST}$.
**Table S4.** Influence of migration probability ($\lambda$) on the sensitivity, specificity and variance of $G_{min}$ and $F_{ST}$.
**Fig. S1.** A) Sensitivity of the $F_{ST}$ and $G_{min}$ measures for varying rates of migration (migration probability) and time of migration (relative to time of population divergence). The left column shows plots of sensitivity for $F_{ST}$ and the right column shows sensitivity for $G_{min}$. The top row shows sensitivity when a significance threshold of 5% is used, the middle row shows a threshold of 1%, and the bottom row shows a threshold 0.1%. B) Specificity of the $F_{ST}$ and $G_{min}$ measures for varying rates and times of migration. Layout of the plots are the same as in panel A.
**Fig. S2.** Sensitivity of $F_{ST}$ (left column) and $G_{min}$ (right column) for varying levels of population recombination rate: $\rho = 0$ (top), $\rho = 50$ (middle), and $\rho = 150$ (bottom).
**Fig. S3.** Specificity of $F_{ST}$ (left column) and $G_{min}$ (right column) for varying levels of population recombination rate: $\rho = 0$ (top), $\rho = 50$ (middle), and $\rho = 150$ (bottom).
**Fig. S4.** Sensitivity of $F_{ST}$ (left column) and $G_{min}$ (right column) for varying levels of population mutation rate: $\theta = 10$ (top), $\theta = 50$ (middle), and $\theta = 150$ (bottom).
**Fig. S5.** Specificity of $F_{ST}$ (left column) and $G_{min}$ (right column) for varying levels of population mutation rate: $\theta = 10$ (top), $\theta = 50$ (middle), and $\theta = 150$ (bottom).
**Fig. S6.** Sensitivity of $F_{ST}$ (left column) and $G_{min}$ (right column) for varying sample size: $n_2 = 10$ (top) and $n_2 = 1$ (bottom).
**Fig. S7.** Specificity of $F_{ST}$ (left column) and $G_{min}$ (right column) for varying sample sizes: $n_2 = 10$ (top) and $n_2 = 1$ (bottom).
**Fig. S8.** Neighbor-joining trees showing the first newly identified region of gene flow on the *Drosophila melanogaster* X chromosome between coordinates 1.65-2.05 Mb. 12.95-13 Mb
**Fig. S9.** Neighbor-joining trees showing the second newly identified region of gene flow on the *Drosophila melanogaster* X chromosome between coordinates 12.95-13 Mb.
**Fig. S10.** $G_{min}$ and $F_{ST}$ scans of the *Drosophila melanogaster* X chromosome in differently sized windows.

**Table 1.** Variance partitioning for $G_{min}$ and $F_{ST}$ under the isolation model of divergence. Column values are the percent variance in each of the two statistics and their respective standard deviations (SD) described by each model parameter (or interaction of parameters), including the population divergence time $\tau_D$, the population mutation ($\theta$) and recombination ($\rho$) rates, and the sample size from a second population ($n_2$). The table only includes parameters with an effect greater than 1%.

| Parameter | $G_{min}$ | SD($G_{min}$) | $F_{ST}$ | SD($F_{ST}$) |
|---|---|---|---|---|
| $\tau_D$ | 48.5 | 44.9 | 54.7 | 27.3 |
| $\theta$ | 3.9 | 2.4 | 0 | 1.4 |
| $n_2$ | 7.6 | 0.1 | 29.3 | 5.5 |
| $\rho$ | 2.5 | 16 | 0 | 39.1 |
| $\tau_D \times n_2$ | 2.9 | 1.3 | 5.4 | 0.2 |
| $\tau_D \times \rho$ | 3.9 | 5.3 | 0 | 2.1 |
| Coalescent processes | 29.8 | 28.8 | 10.5 | 23.2 |

**Figure legends**

**Fig. 1.** Examples of the average and minimum between population coalescent times for models that include A) population divergence in isolation, and B) secondary contact. For sufficiently high rates of mutation, these two times are the main determinants for the observable quantities: the mean number of between population nucleotide differences, $\bar{d}_{xy}$, and the minimum between population differences, $\min(d_{xy})$.

**Fig. 2.** A) The mean simulated values of $G_{min}$ plotted against divergence time for a model of divergence in isolation. B) Mean simulated values of $G_{min}$ plotted against population mutation rate under an isolation model with divergence occurring at time $\tau_D = N$ generations ago. C) and D) The mean simulated values of $F_{ST}$ plotted against divergence time (C) and population mutation rate (D) under an isolation model. The shaded areas delimit the mean ± one standard deviation. The blue lines represent sample sizes in the two populations of $n_1 = 10$ and $n_2 = 10$, while the red lines represent sample sizes of $n_1 = 10$ and $n_2 = 1$. The simulations shown here do not include the effects of intra-locus recombination.

**Fig. 3.** Heatmaps of percent improvement of $G_{min}$ over $F_{ST}$ for sensitivity (left) and specificity (right). Improvement was calculated for varying rates of migration (migration probability) and time of migration (relative to time of population divergence) and averaged over all other parameters.

**Fig. 4.** Cosmopolitan admixture in sub-Saharan African *Drosophila melanogaster*. A) $G_{min}$ (above) and $F_{ST}$ (below) in 50 kb windows across the X chromosome in a sample of seven French and nine Rwandan lines. Shaded regions indicate where Pool *et al.* (2012) previously detected admixture. Open circles mark windows that are identified as outliers from the chromosome-wide distribution. B) Scatterplot of $F_{ST}$ versus $G_{min}$ across the X chromosome in 50 kb windows. The diagonal line was added for reference.

Figure 1

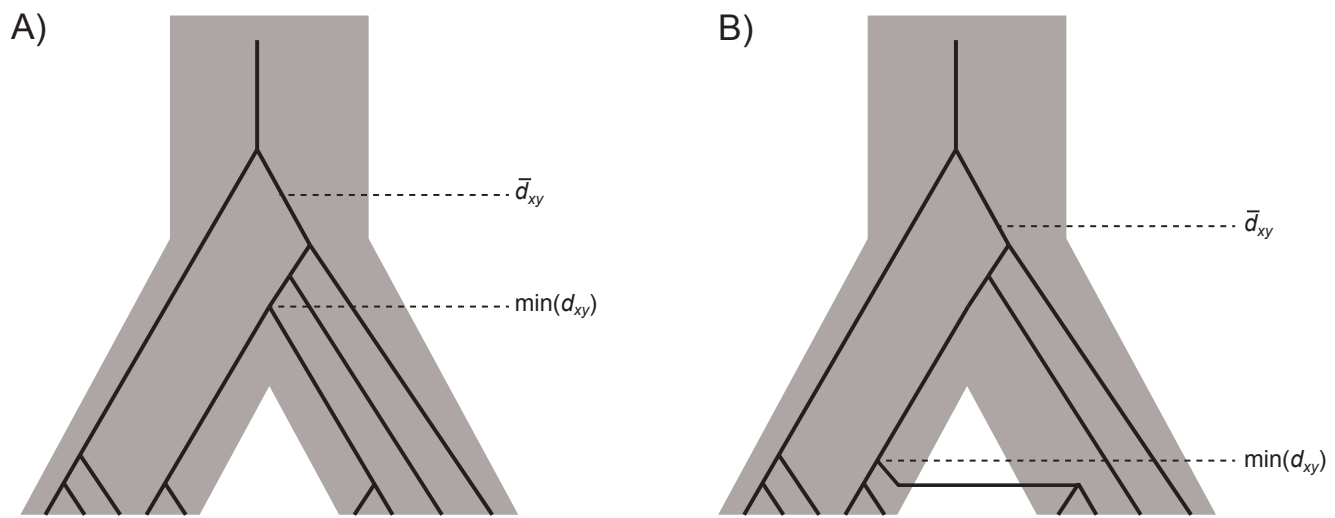

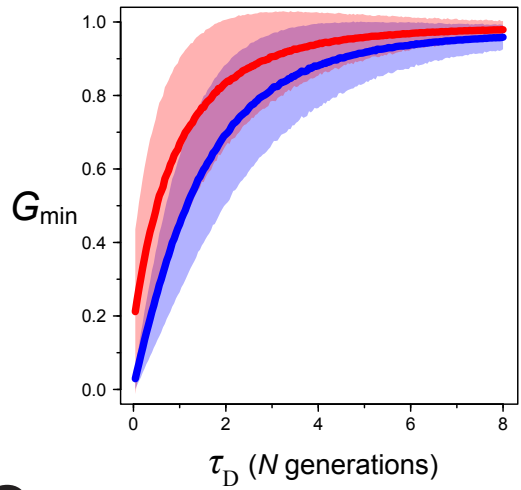
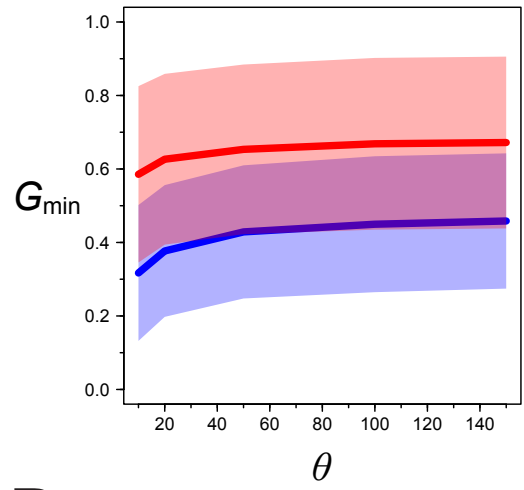
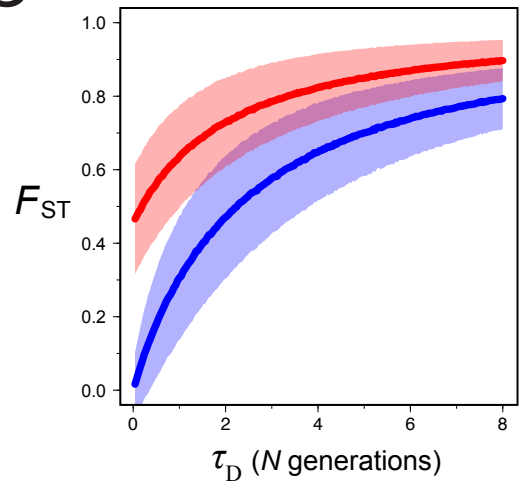
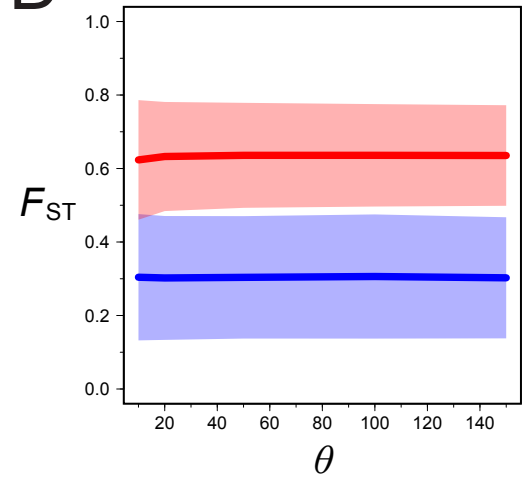

Figure 3

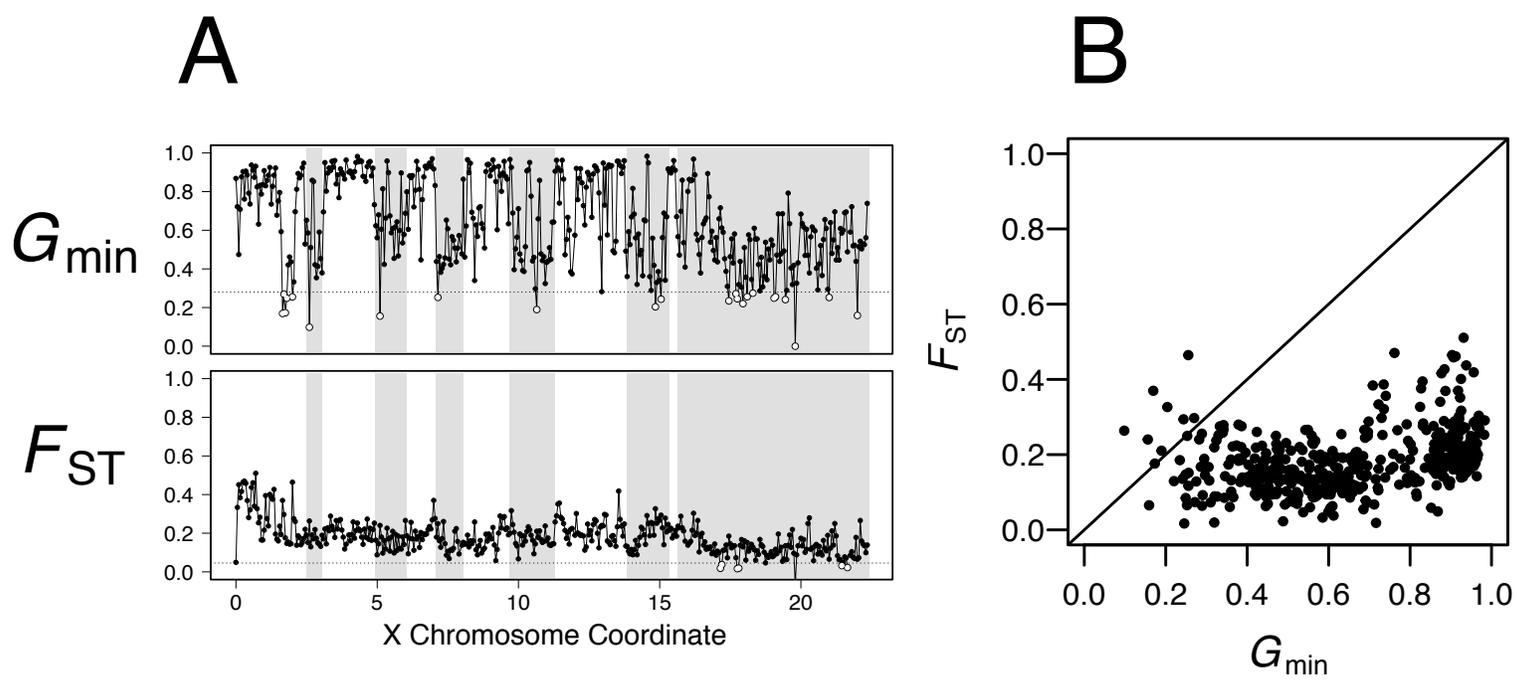

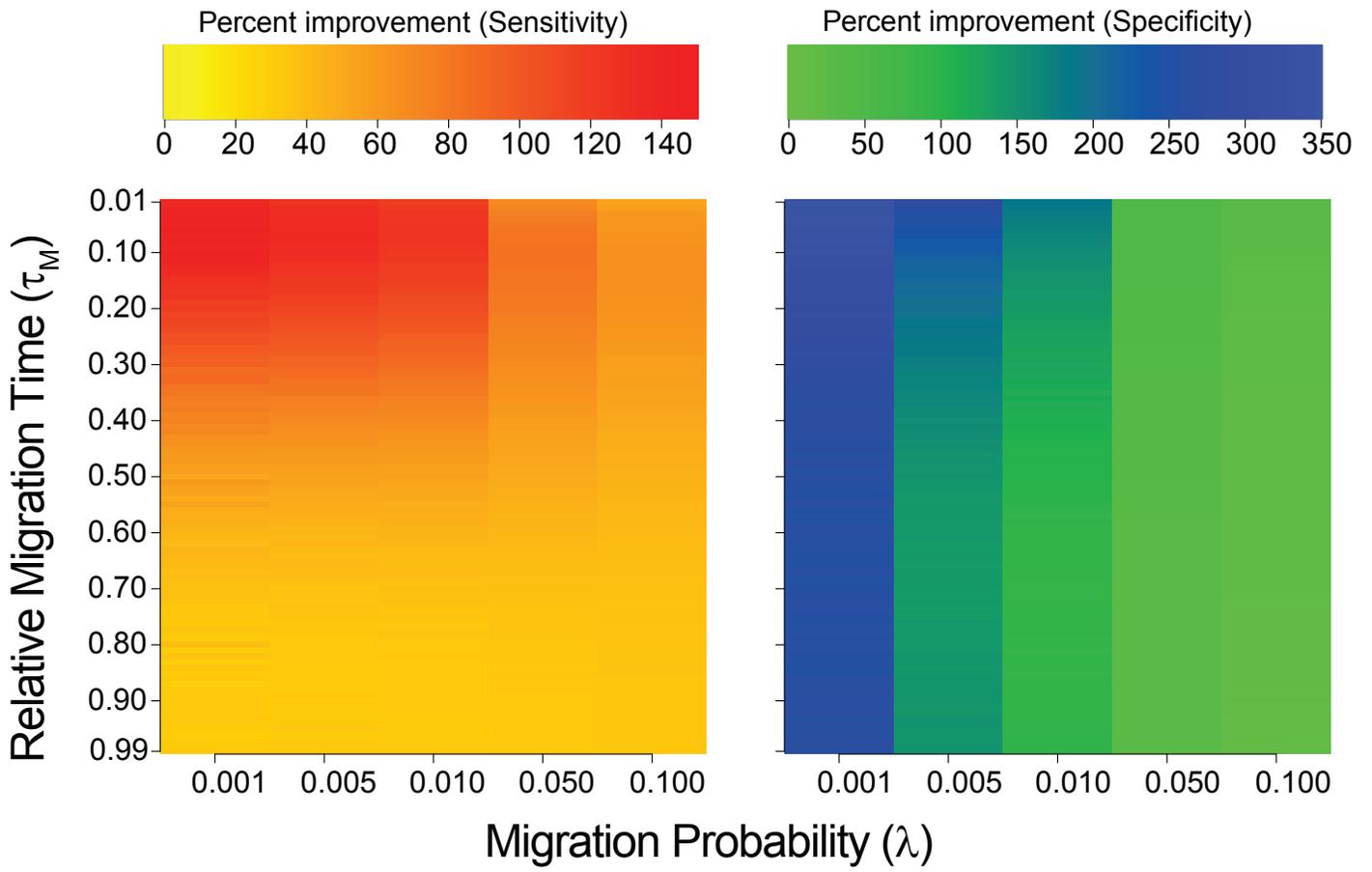

**Table S1.** The sampled strains from two populations of *Drosophila melanogaster* used to contrast the performance of $F_{ST}$ and $G_{min}$, including short read archive accession numbers.

| Population | Stock | Accession |
|---|---|---|
| France | FR14 | SRR189088 |
| France | FR151 | SRR189089 |
| France | FR207 | SRR189091 |
| France | FR217 | SRR189092 |
| France | FR229 | SRR189093 |
| France | FR310 | SRR189094 |
| France | FR361 | SRR189095 |
| Rwanda | RG10 | SRR189374 |
| Rwanda | RG11N | SRR189375 |
| Rwanda | RG15 | SRR189377 |
| Rwanda | RG15 | SRR189392 |
| Rwanda | RG21N | SRR189382 |
| Rwanda | RG32N | SRR189388 |
| Rwanda | RG34 | SRR189390 |
| Rwanda | RG35 | SRR189391 |
| Rwanda | RG35 | SRR189378 |
| Rwanda | RG36 | SRR189393 |
| Rwanda | RG38N | SRR189395 |

**Table S2.** Analysis of variance of sensitivity of $G_{min}$ and $F_{ST}$ measures at varying quantile thresholds under a model including gene flow. Column values are the percent variance in each measure described by each parameter or interaction of parameters, filtered to include only those with an effect greater than 1%.

| Parameter | 5% quantile | | 1% quantile | | 0.1% quantile | |
|---|---|---|---|---|---|---|
| | $G_{min}$ | $F_{ST}$ | $G_{min}$ | $F_{ST}$ | $G_{min}$ | $F_{ST}$ |
| $\tau_D$ | 9.7 | 6.8 | 4.3 | 3.3 | 2.3 | 1.6 |
| $\rho$ | 0.0 | 0.3 | 1.2 | 0.2 | 1.2 | 0.4 |
| $n_2$ | 7.0 | 25.9 | 4.3 | 14.1 | 2.6 | 3.9 |
| $\lambda$ | 1.4 | 0.3 | 1.1 | 0.2 | 0.8 | 0.2 |
| $\tau_M$ | 23.9 | 13.3 | 16.7 | 9.3 | 11.8 | 4.4 |
| $\tau_D \times \rho$ | 4.3 | 0.5 | 1.0 | 0.1 | 0.2 | 0.0 |
| $\tau_D \times n_2$ | 0.5 | 0.8 | 0.4 | 2.3 | 0.4 | 1.5 |
| $\rho \times n_2$ | 1.1 | 1.5 | 0.1 | 0.0 | 0.0 | 0.4 |
| $\tau_D \times \tau_M$ | 5.2 | 3.7 | 4.1 | 3.9 | 3.2 | 2.7 |
| $n_2 \times \tau_M$ | 3.6 | 9.1 | 3.7 | 8.6 | 3.1 | 4.4 |
| $\lambda \times \tau_M$ | 1.8 | 0.3 | 1.7 | 0.3 | 1.5 | 0.3 |
| $\tau_D \times \rho \times \tau_M$ | 1.9 | 0.2 | 0.8 | 0.0 | 0.2 | 0.1 |
| $\tau_D \times n_2 \times \tau_M$ | 0.3 | 2.4 | 0.6 | 3.4 | 0.6 | 2.6 |
| coalescent | 33.0 | 32.7 | 56.0 | 51.8 | 67.3 | 73.3 |

**Table S3.** Analysis of variance of specificity of the $F_{ST}$ and $G_{min}$ measures at varying quantile thresholds under a model including gene flow. Column values are the percent variance in each measure described by each parameter or interaction of parameters, filtered to include only those with an effect greater than 1%.

| Parameter | 5% quantile | | 1% quantile | | 0.1% quantile | |
| --- | --- | --- | --- | --- | --- | --- |
| | $G_{min}$ | $F_{ST}$ | $G_{min}$ | $F_{ST}$ | $G_{min}$ | $F_{ST}$ |
| $\tau_D$ | 0.4 | 1.5 | 1.1 | 5.2 | 1.2 | 10.8 |
| $\Theta$ | 0.7 | 0.1 | 0.8 | 0.6 | 1.1 | 5.2 |
| $\rho$ | 1.2 | 5.8 | 0.7 | 7.7 | 0.2 | 0.8 |
| $n_2$ | 14.9 | 13.3 | 11.6 | 9.7 | 10.3 | 1.9 |
| $\lambda$ | 47.1 | 54.1 | 48.6 | 43.8 | 46.6 | 32.8 |
| $\tau_M$ | 6.4 | 0.9 | 9.0 | 2.5 | 13.2 | 6.9 |
| $\tau_D \times \lambda$ | 0.2 | 0.9 | 0.1 | 2.4 | 0.1 | 2.2 |
| $\rho \times P_M$ | 1.7 | 2.4 | 1.1 | 3.3 | 0.8 | 1.2 |
| $n_2 \times \lambda$ | 2.7 | 7.3 | 1.2 | 2.5 | 0.3 | 0.0 |
| $\tau_D \times \tau_M$ | 0.3 | 0.3 | 0.7 | 0.8 | 0.3 | 1.5 |
| $\Theta \times \tau_M$ | 0.5 | 0.0 | 0.5 | 0.3 | 0.5 | 1.6 |
| $\rho \times \tau_M$ | 2.3 | 0.3 | 2.2 | 0.3 | 2.1 | 0.9 |
| $n2 \times \tau_M$ | 1.6 | 0.2 | 0.8 | 0.0 | 0.8 | 0.0 |
| $n_2 \times \lambda \times \tau_M$ | 0.2 | 0.0 | 0.6 | 0.1 | 1.0 | 0.1 |
| coalescent | 16.0 | 10.1 | 18.8 | 17.6 | 19.6 | 30.3 |

**Table S4.** Sensitivity, specificity and variance of $F_{ST}$ and $G_{min}$ statistics for varying values of probability of migration ($\lambda$)

| Migration Probability ($\lambda$) | Sensitivity | | Specificity | | Variance | |
|---|---|---|---|---|---|---|
| | $G_{min}$ | $F_{ST}$ | $G_{min}$ | $F_{ST}$ | $G_{min}$ | $F_{ST}$ |
| 0.001 | 0.114 | 0.060 | 0.221 | 0.059 | 0.0230 | 0.0002 |
| 0.005 | 0.111 | 0.060 | 0.293 | 0.109 | 0.0232 | 0.0002 |
| 0.010 | 0.107 | 0.059 | 0.357 | 0.164 | 0.0234 | 0.0002 |
| 0.050 | 0.088 | 0.055 | 0.610 | 0.448 | 0.0246 | 0.0002 |
| 0.100 | 0.076 | 0.051 | 0.743 | 0.614 | 0.0255 | 0.0002 |

Figure S1

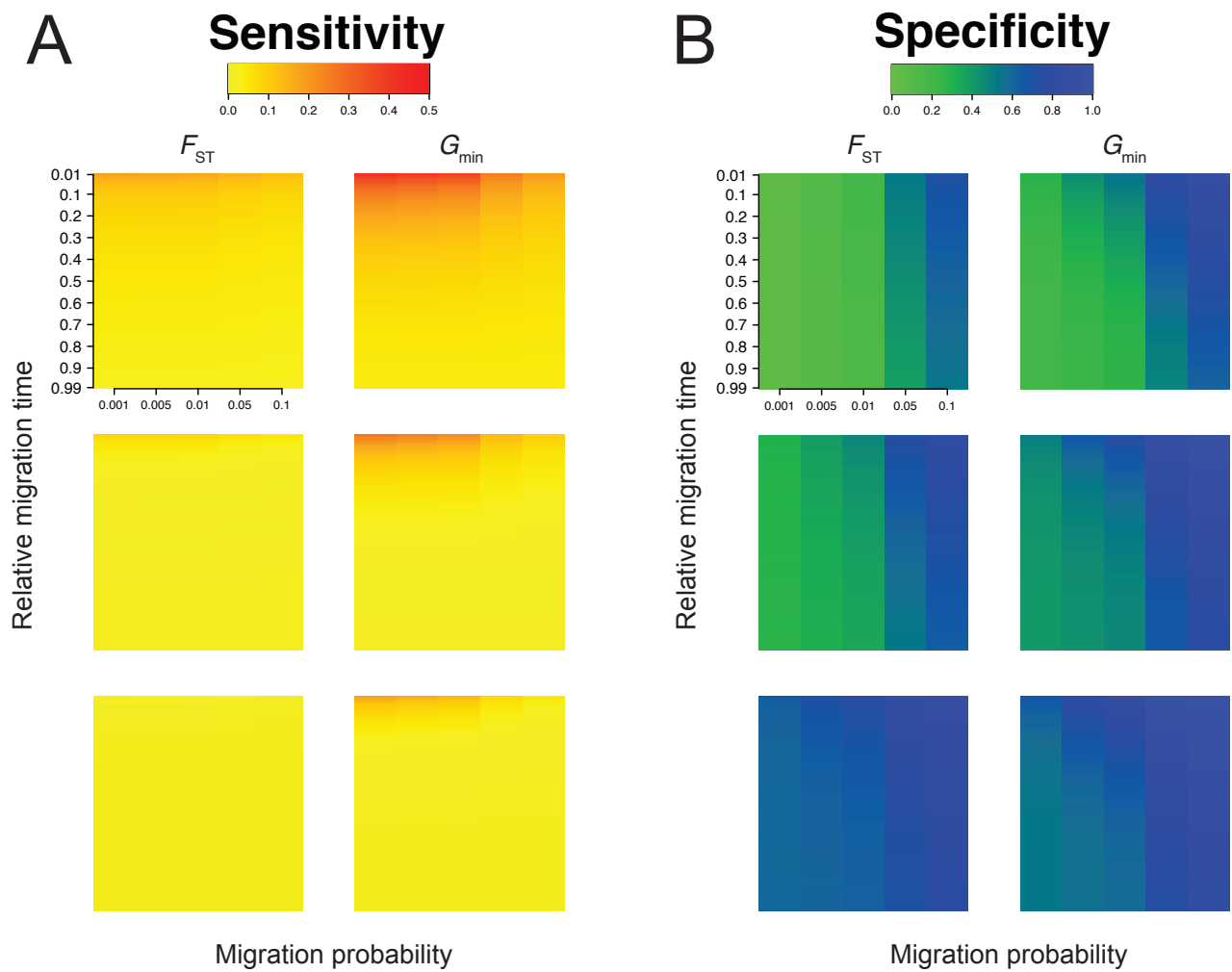

Figure S2

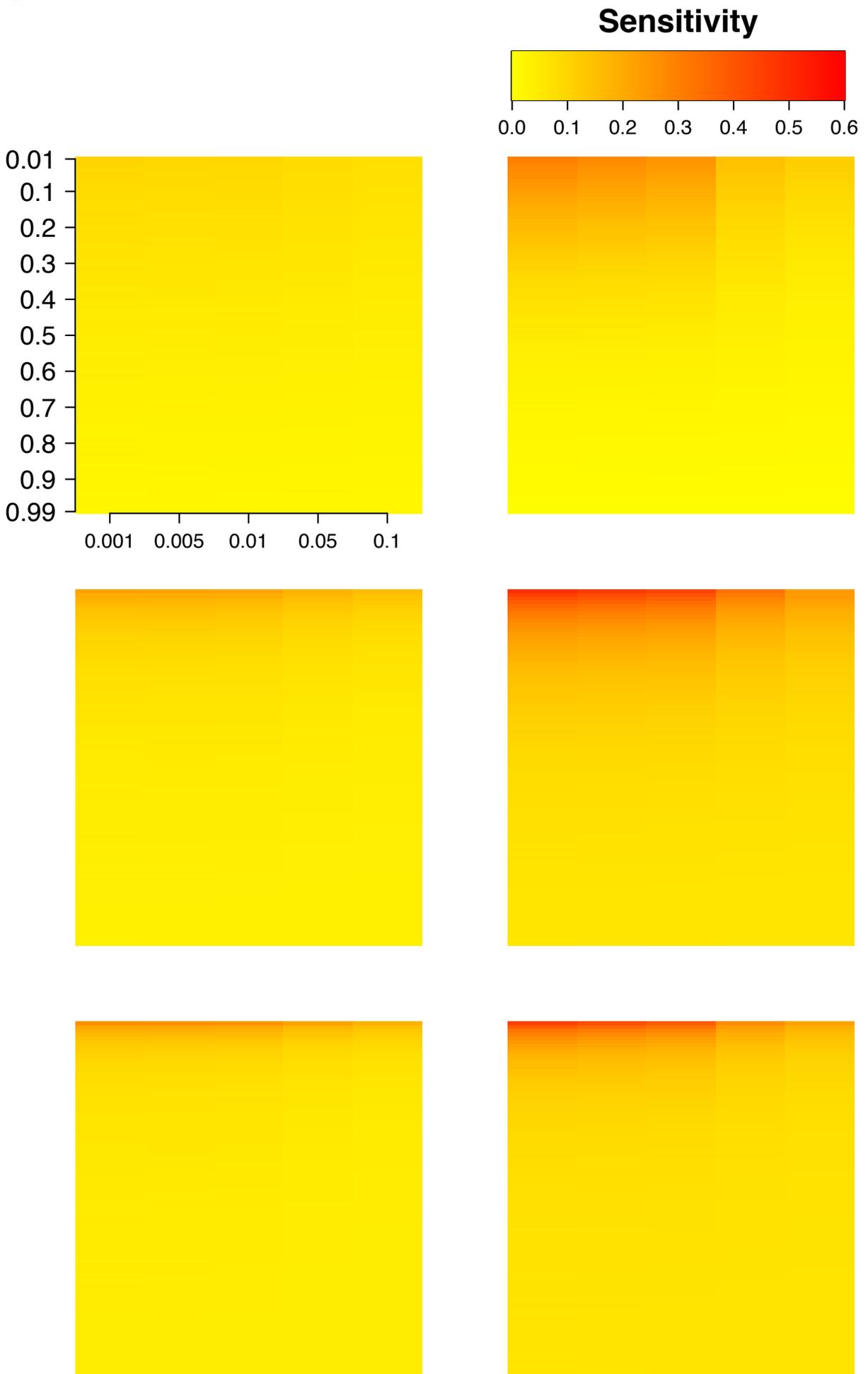

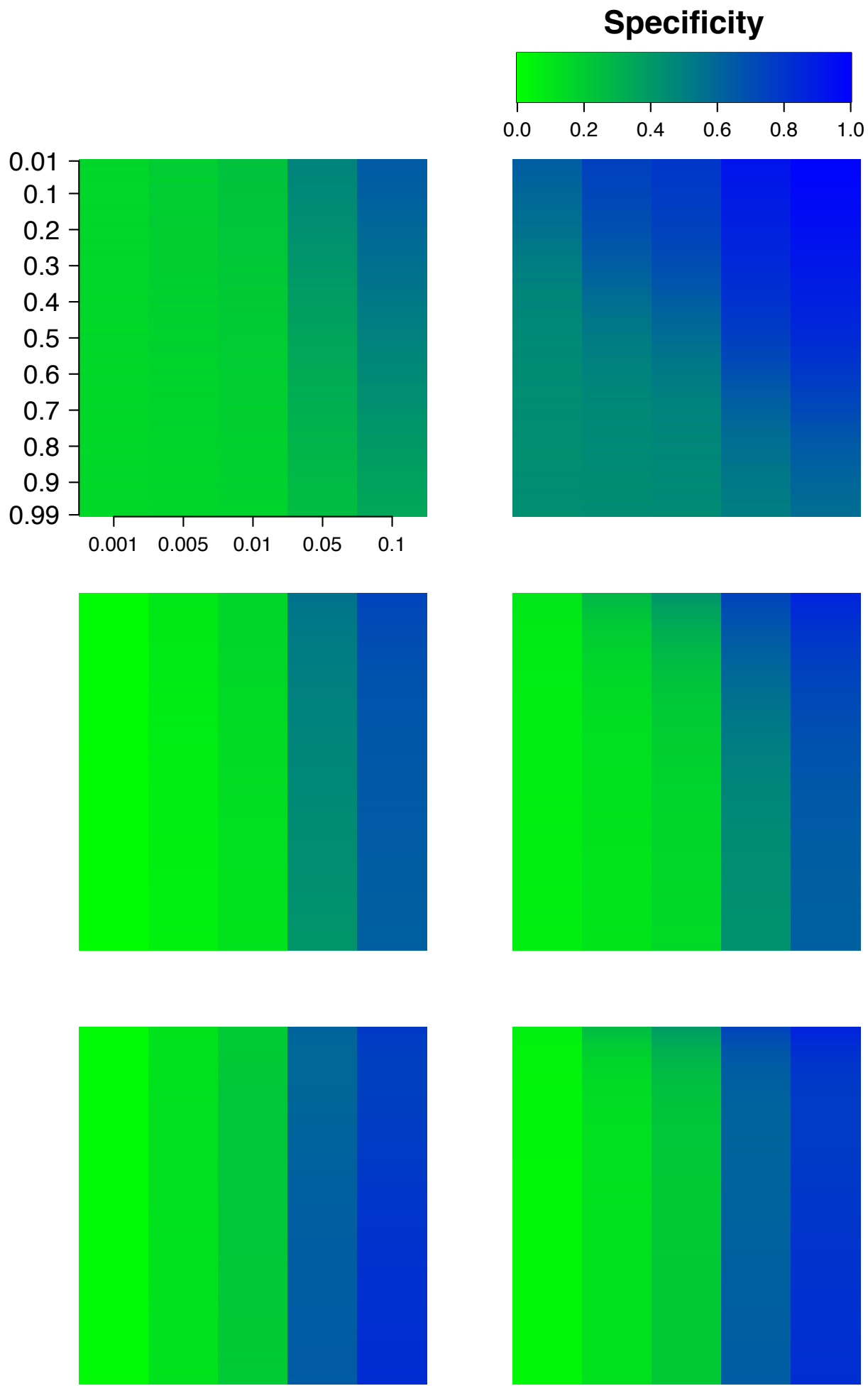

Figure S3

Figure S4

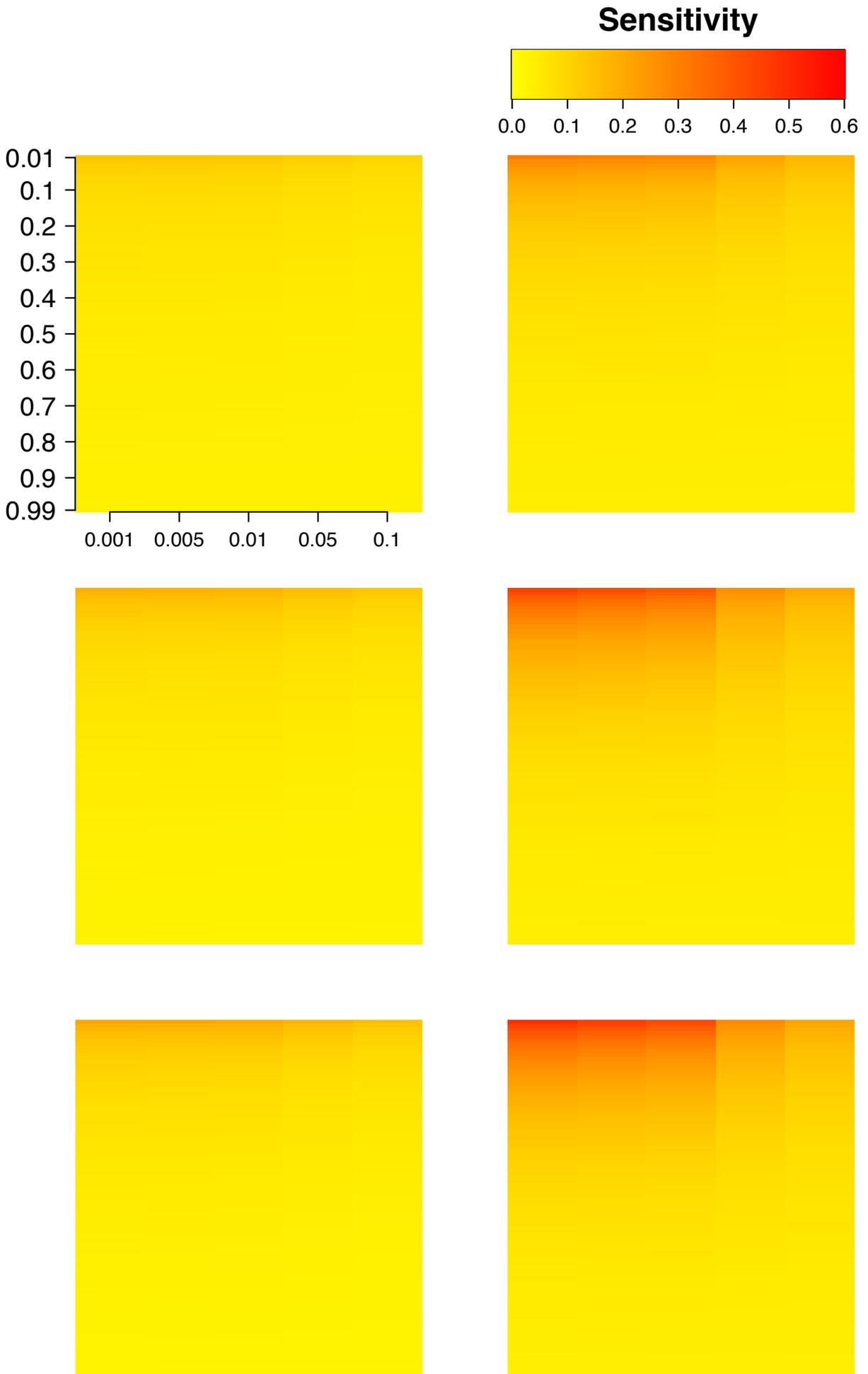

Figure S5

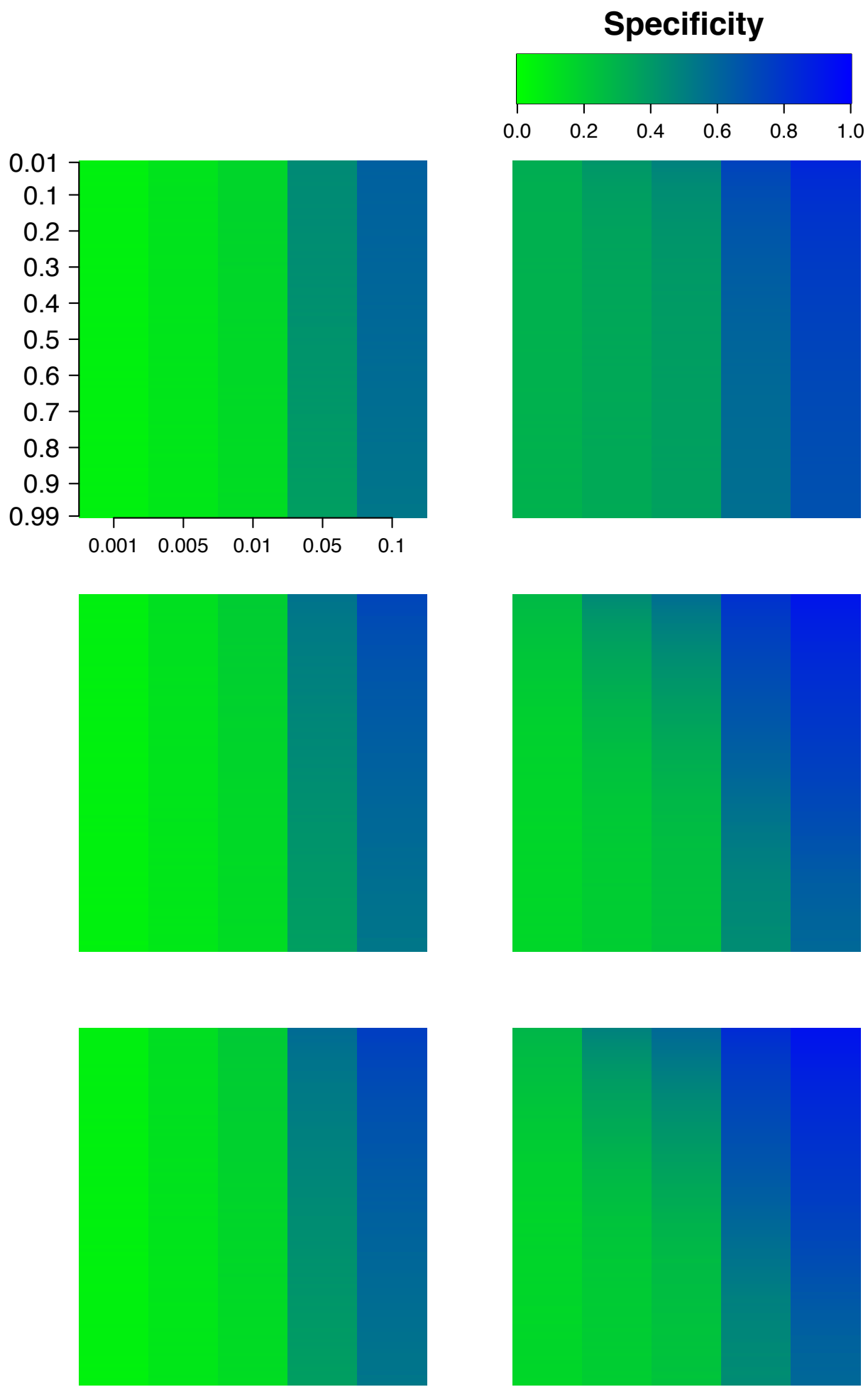

Figure S6

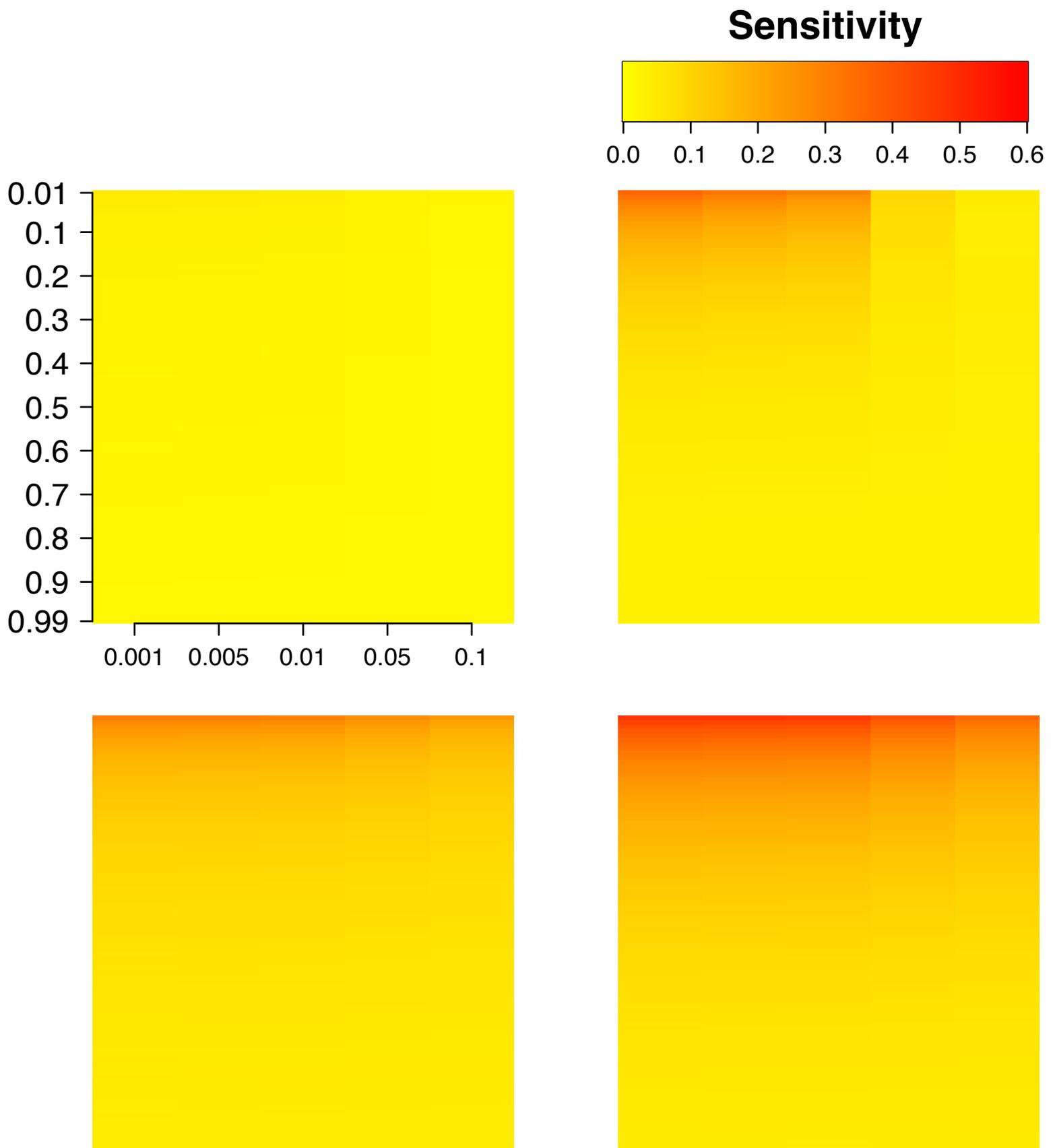

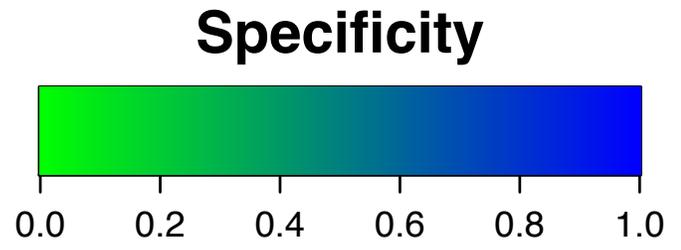
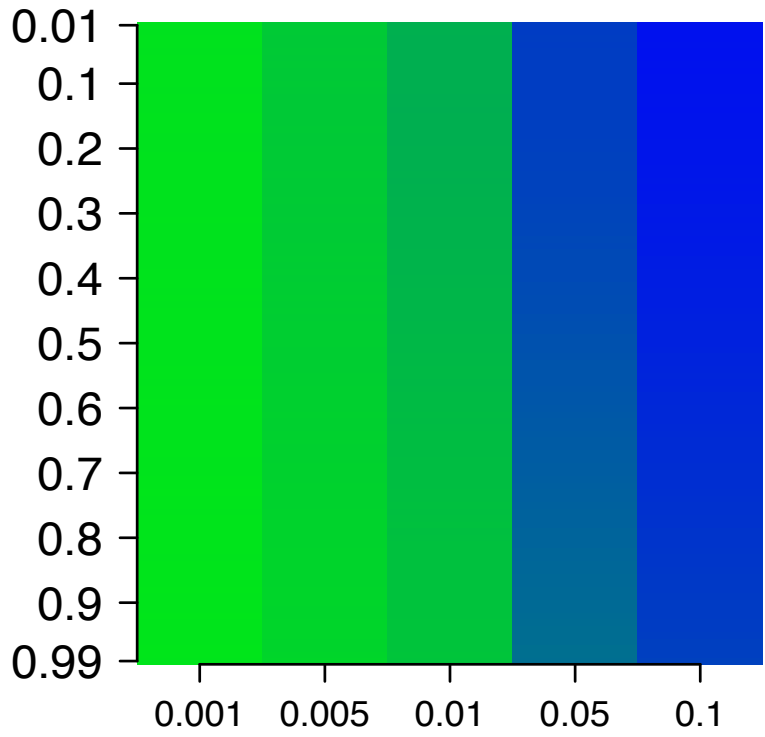
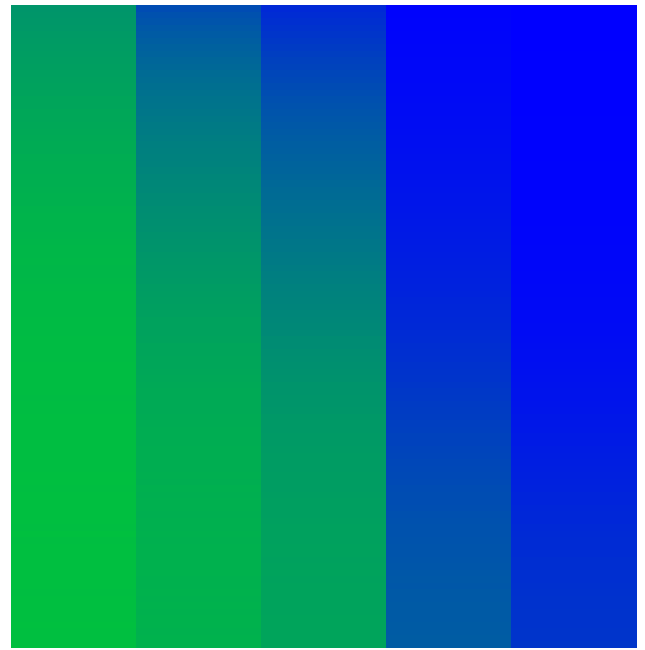
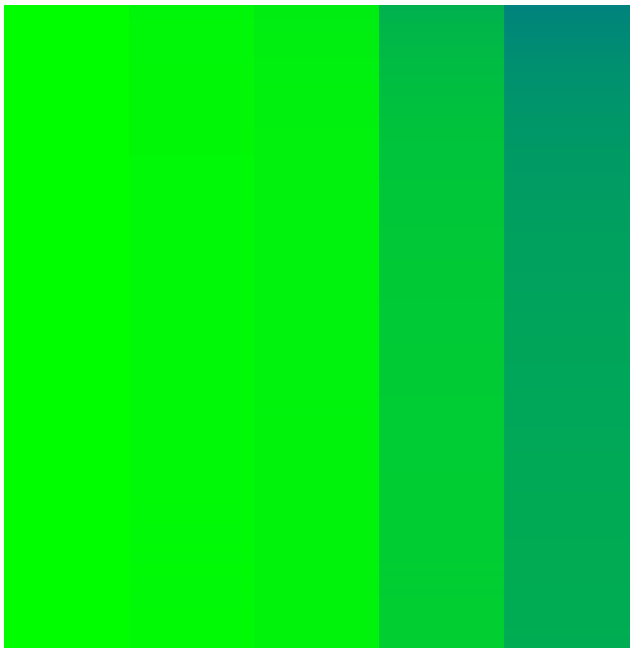
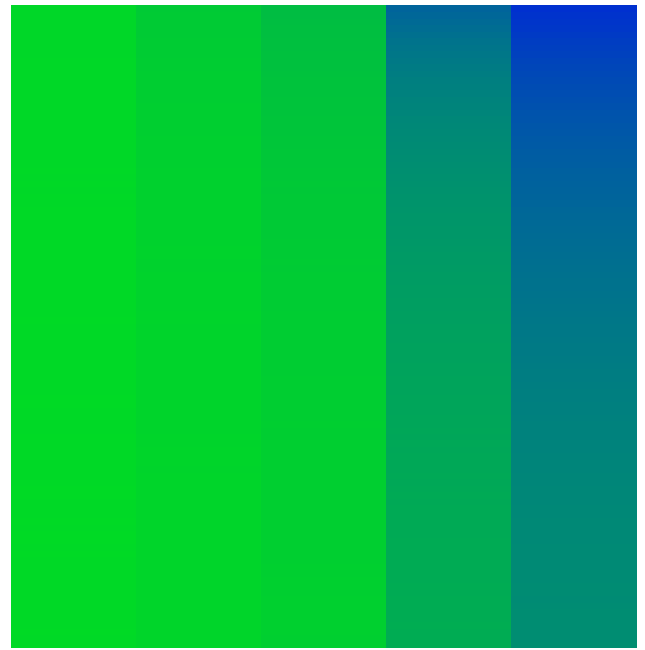

A

B

$G_{min}$

X Chromosome Coordinate

Figure S9

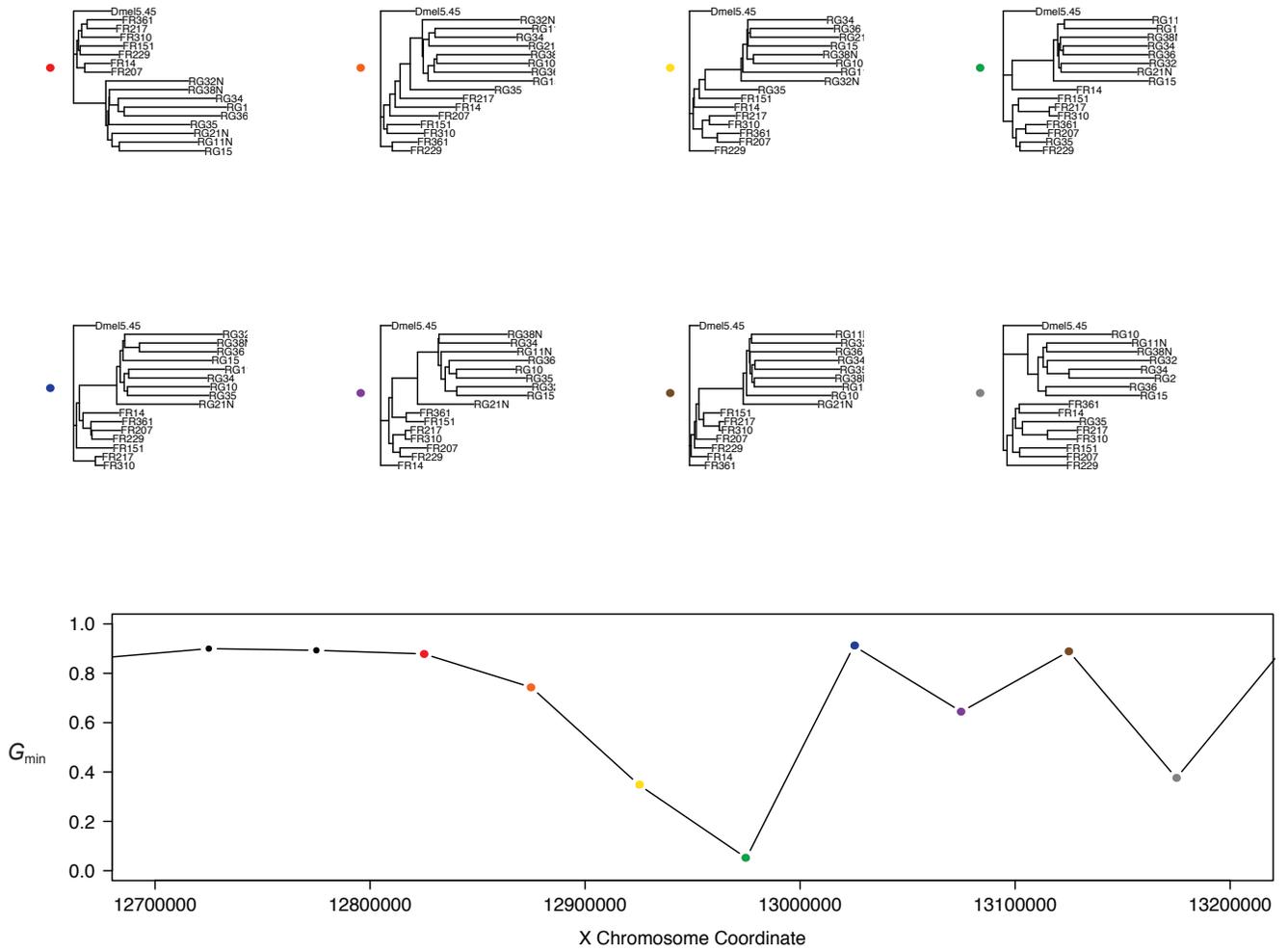

Figure S10

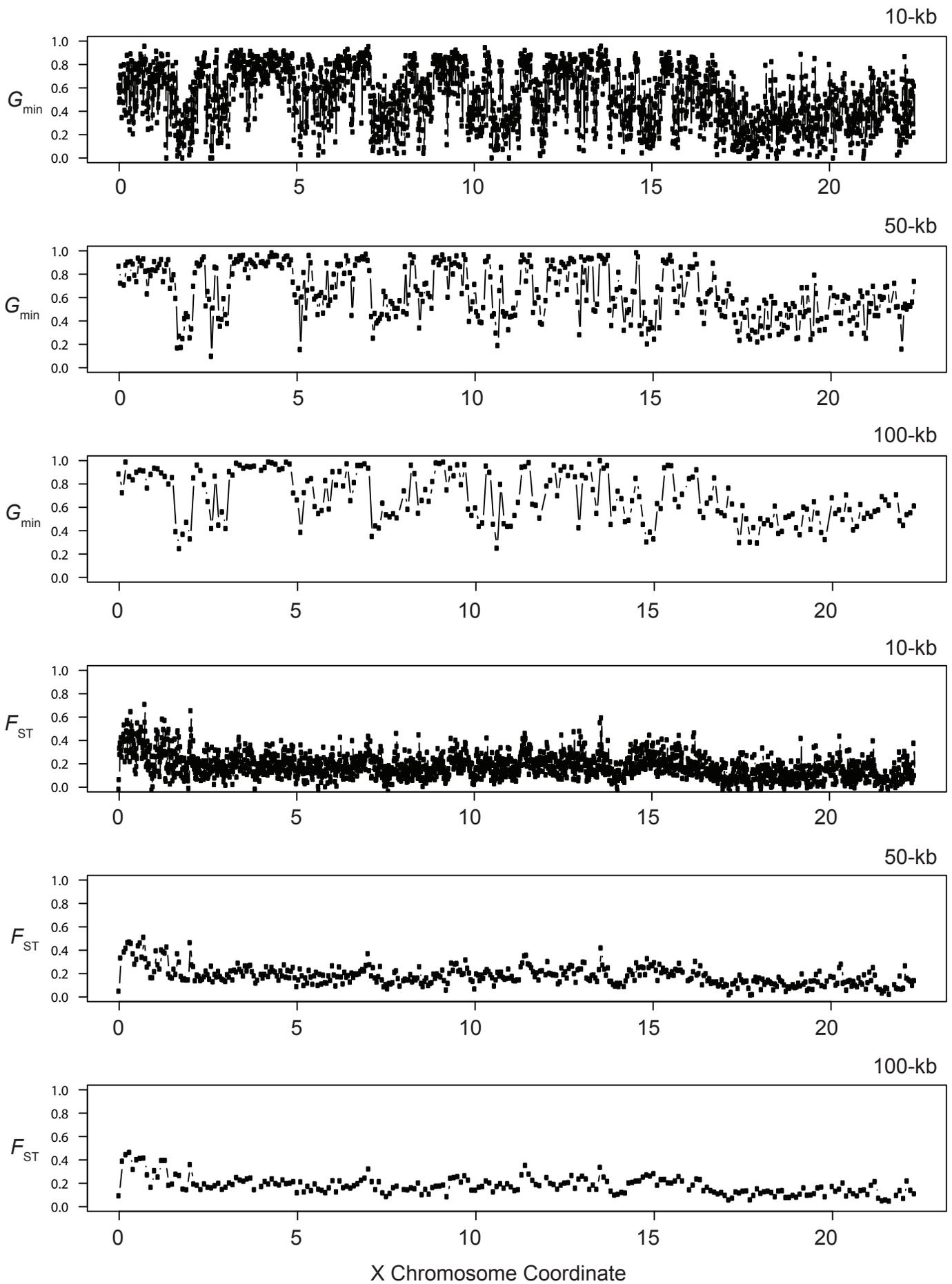